\documentclass[12pt]{article}
\usepackage[a4paper, total={7in, 10in}]{geometry}
\usepackage[parfill]{parskip}
\usepackage{physics, tensor, float, subcaption}
\usepackage{graphicx}
\graphicspath{ {Plots/} }
\usepackage{jhep-mod}
\usepackage{bm}
\usepackage{soul}
\usepackage{amssymb,amsmath,amsthm}
\usepackage{mathrsfs}
\usepackage[utf8]{inputenc}
\usepackage{lmodern} 
\usepackage[newzealand]{babel}
\usepackage[style=american]{csquotes} 
\usepackage{microtype}
\usepackage[T1]{fontenc} 
\usepackage{enumerate}
\usepackage{bigints}
\usepackage{xcolor}
\usepackage{appendix}
\usepackage{graphicx}
\usepackage{float}
\usepackage{tikz}
\usepackage{setspace}
\usepackage{cancel}
\usepackage{doi}
\definecolor{purple}{rgb}{1,0,1}
\definecolor{darkgreen}{rgb}{.125,.5,.25}

\definecolor{lime}{HTML}{A6CE39}
\newcommand{\orcidicon}{%
	\begin{tikzpicture}
	\draw[lime, fill=lime] (0,0) 
		circle [radius=0.16] 
		node[white] {{\fontfamily{qag}\selectfont \tiny ID}};
	\draw[white, fill=white] (-0.0625,0.095) 
		circle [radius=0.007];
	\end{tikzpicture}
	\hspace{-5mm}
}
\newcommand\orcidJessica{{\href{https://orcid.org/0000-0002-2669-2899}{\orcidicon}}}
\newcommand\orcidSebastian{{\href{https://orcid.org/0000-0003-1997-0026}{\orcidicon}}}
\newcommand\orcidMatt{{\href{https://orcid.org/0000-0003-1088-6485}{\orcidicon}}}
\begin{document}

\title{\vspace{-125pt}\huge{
ADM mass in warp drive spacetimes
}}

\author{
\Large

Sebastian Schuster\!\orcidSebastian$^1$,
Jessica Santiago\!\orcidJessica$^2$,\\  {\sf  and} Matt Visser\!\orcidMatt$^3$}
\affiliation{$^1$ Institute of Theoretical Physics,
Faculty of Mathematics and Physics, \\
\null\qquad Charles University,
 V~Hole\v{s}ovi\v{c}k\'{a}ch~2,
180 00 Prague 8,
Czech Republic.}
\affiliation{$^2$ Section of Astrophysics, Astronomy and Mechanics, 
	Department of Physics,\\
	\null\qquad 
	Aristotle University of Thessaloniki, Thessaloniki 54124, Greece.}

\affiliation{$^3$ School of Mathematics and Statistics, Victoria University of Wellington, 
	\\
	\null\qquad PO Box 600, Wellington 6140, New Zealand.}
\emailAdd{sebastian.schuster@utf.mff.cuni.cz}
\emailAdd{jessicasantiago@auth.gr}
\emailAdd{matt.visser@sms.vuw.ac.nz}

\abstract{
What happens when a warp bubble has mass? 
This seemingly innocent question forces one to carefully formalize exactly what one means by a warp bubble, exactly what one means by having the warp bubble ``move'' with respect to the fixed stars, and forces one to more carefully examine the notion of mass in warp-drive spacetimes. This is the goal of the present article. In this process, we will see that often-made throw-away comments regarding ``payloads'' are even simpler than commonly assumed, while there are two further, distinct yet subtle ways in which a mass can appear in connection with a warp drive space-time: One, that the warp bubble (not its payload) has the mass; two, that the mass is a background feature in front of which the warp drive moves. For simplicity, we consider generic Natário warp drives with zero-vorticity flow field. The resulting spacetimes are sufficiently simple to allow an \emph{exact} and \emph{fully explicit} computation of \emph{all} of the stress-energy components, and verify that (as expected) the null energy condition (NEC) is violated. Likewise the weak, strong, and dominant energy conditions (WEC, SEC, DEC) are violated. 
Indeed, this confirms the community's folk wisdom, and recent (fully general, but implicit) results of the present authors which closed previous gaps in the argument. However, folk wisdom should be carefully and critically examined before being believed, and the present examples for general results will greatly aid physical intuition.

\bigskip
\noindent
{\sc Date:} Wednesday 1 June 2022;  4 August 2022; \LaTeX-ed \today


}

\maketitle
\def\tr{{\mathrm{tr}}}
\def\diag{{\mathrm{diag}}}
\def\cof{{\mathrm{cof}}}
\def\pdet{{\mathrm{pdet}}}
\def\d{{\mathrm{d}}}
\def\K{{\mathcal{K}}}
\def\O{{\mathcal{O}}}
\parindent0pt
\parskip7pt
\newcommand{\C}{\mathcal{C}}

\clearpage

\section{Introduction}
\label{S:intro}

Warp bubbles and warp drives by now have a quite extensive 27-year history in the general relativity community. First formulated by Alcubierre in 1994~\cite{Alcubierre:1994},  two distinct  modifications (zero-expansion and generic) were subsequently developed by Natário~\cite{Natario:2001} in 2001. Considerable work along these lines has continued~\cite{Everett:1995, Everett:1997, Hiscock:1997, Pfenning:1998, Low:1998, Olum:1998, VanDenBroeck:1999a, VanDenBroeck:1999b, Clark:1999,
Visser:1999, Alcubierre:2001, Lobo:2002, Lobo:2004a, Lobo:2004b, Lobo:2007, Finazzi:2009, McMonigal:2012, Lobo:2017, Alcubierre:2017},
largely focussing on the Alcubierre and Natário zero-expansion variants, 
culminating in a recent focus on zero-vorticity warp drives~\cite{Lentz:2020, Lentz:2021,  Fell:2021, Santiago:2021a},
and the closely related, and very recently introduced notions of tractor/pressor/stressor beams \cite{Santiago:2021b, Visser:2021}.

A mainstay in warp drive research past and present has been the study of energy conditions in such space-times, just as in many other reverse-engineered space-time geometries such as wormholes, the Gödel universe, or Krasnikov hypertubes. Specifically, this last year has seen multiple misguided claims~\cite{Lentz:2020, Lentz:2021, Bobrick:2021, Fell:2021, Osvaldo-1, Osvaldo-2, Osvaldo-3, Osvaldo-4, Osvaldo-5} as to the asserted possible avoidance/amelioration of the known energy condition violations in warp drive spacetimes; violations that were first established over 20 years ago, to varying degrees of generality.
We have previously provided~\cite{Santiago:2021a}, for the first time, a full proof of NEC violations in all generic Natário warp drive spacetimes (using implicit arguments in the form of a proof by contradiction), thereby firmly establishing that those claims of \emph{non}-violation are irretrievably erroneous. In the current article we will focus on the zero-vorticity case of Natário warp drives to provide a particularly simple, straightforward and fully explicit calculation showing exactly where the key difficulty lies. In counterpoint we should specifically mention that energy condition violations are not the absolute prohibition that they have often been taken to be in the past --- energy condition violations are instead an invitation to think very carefully about the underlying physics~\cite{Santiago:2021a,Santiago:2021b,Visser:2021, twilight, Curiel, Kontou}. 
\enlargethispage{20pt}

Additionally, and possibly more importantly than yet-another demonstration of violated energy conditions, this allows for the first time an explicit discussion of the meaning of mass in warp drive space-times. We start by asking the innocent sounding question ``What happens when one puts a Schwarzschild black hole into a warp bubble?''.
Answering this question forces one to focus on 
several basic and fundamental issues: What does one mean by a moving warp bubble? How does one formalize the notion of mass for a warp bubble?

\enlargethispage{20pt}
In order to answer these questions, we shall first set up the general framework (making use of the asymptotic flatness of generic Natário warp drives), including a discussion of three distinct notions of mass in a warp bubble context. (That the notion of mass in any general relativistic context is a subtle one might be gleaned from references~\cite{Wang:2020,Szabados:2009,Cederbaum}, and the textbook discussions in~\cite[chapter 19]{MTW} and~\cite[chapter 11.2]{Wald}.)
We then specialize to zero-vorticity warp bubbles, (which are closely related to the Painlev\'e--Gullstrand version of Schwarzschild spacetime), and carefully distinguish a moving zero-vorticity warp bubble from, for instance,  the Painlev\'e--Gullstrand version of Schwarzschild spacetime. 

\clearpage

This allows us to  define a precise notion of motion (of the warp bubble with respect to the fixed stars), and explicitly calculate the stress-energy tensor. Within this zero-vorticity framework we then first develop a 
Schwarzschild-based warp drive --- wherein a Schwarzschild black hole is embedded in a warp bubble, and then develop a more specialized framework suitable for describing a finite-mass payload (spacecraft) embedded in a warp bubble.
The zero-vorticity case is sufficiently simple to allow fully explicit computation of the stress-energy, and explicitly verify violations of the NEC. In this regard the zero-vorticity warp bubble closely follows the fully explicit computations that have previously been carried out for the Alcubierre and zero-expansion warp bubbles.

The article is organized as follows: In section~\ref{S:basics}, we introduce the notation used in this paper, and the geometry and basic facts of the generic Natário warp drive. We also explain the three distinct physical ways in which a warp drive spacetime can be combined with relativistic notions of mass. Two of these are investigated subsequently. Section~\ref{S:zero-vorticity} then gives the special zero-vorticity case of the generic Natário warp drive on which we will build our analysis. In section~\ref{S:black-warp}, we define a warped Schwarzschild spacetime,  colloquially referred to as a \enquote{black-warp spacetime}, a Schwarzschild black hole engineered to move through spacetime in a prescribed manner, and then investigate its properties.
Subsequently section~\ref{S:regular} replaces the central black hole by a regular finite-mass payload inside the warp bubble. In section~\ref{S:conclusions}, we conclude and discuss future directions.

\section{Natário generic warp drives}
\label{S:basics}

Let us first discuss generic Natário class of warp drives~\cite{Natario:2001}. This generic Natário class
is broad enough to cover almost all of the relevant literature. 

\subsection{Metric, co-tetrad, and tetrad}

The Natário generic warp bubbles all have line elements explicitly of the form~\cite{Natario:2001}:
\begin{equation}
	\label{E:generic}
	\d s^2 = -\d t^2 + \delta_{ij} \, 
	(\d x^i - v^i(t,\vec x) \d t)\, (\d x^j - v^j(t,\vec x) \d t). 
\end{equation}
The spatial 3-slices are flat Euclidean space; the lapse function is unity; all of the physics is encoded in the flow vector $v^i(t,\vec x)$, which is the negative of what is usually called the shift vector in the ADM formalism~\cite{ADM1,ADM2}. Furthermore, this space-time is assumed to be asymptotically flat. Originally/Traditionally warp drives are subject to the added assumption of global hyperbolicity, but we think this is a misleading property to assume, as many of the more problematic features of warp drives space-times can then not be properly discussed\footnote{And we would like to caution the reader that this assumption might possibly invite proponents of faulty logic, who then conclude the absence of causality issues in superluminal warp drives based on its assumed global hyperbolicity. Much more serious physics issues than the energy condition can occur and need to be addressed in warp drive space-times. A minimal definition ensures easy access to these discussions.}. 
	
Also note that this set-up certainly includes more than just warp drives: Schwarzschild in Painlevé--Gullstrand coordinates is contained in this form. This will play a role in what is to come in section~\ref{S:black-warp}. Also, the static patch of the de Sitter universe can in suitable coordinates be cast in this form, as can all spatially flat FLRW spacetimes~\cite{GaurVisser:2022}. 
Neither of these examples can or should be considered a warp drive. 
Since ---apart from warp drives themselves--- it is clear that other space-times (or parts of space-times) can fit into this general Painlev\'e--Gullstrand form,  it is important to raise the question: How can we safely tell apart a warp drive from a non-warp metric in unusual coordinates? As a first step in order to answer this question, an additional 
interpretative dance around the flow vector $v^i(t,\vec x)$ has to be undertaken.

  Usually, this involves giving the flow vector  a ``Newtonian'' interpretation, whatever this means in general relativity, and wilfully ignoring the fact that the previous non-warp-drive examples allow this interpretation just as well. Still (and sadly?) this is needed to make the very important distinction between superluminal and subluminal warp drives ($v^i v_i > 1$ and $v^i v_i < 1$, respectively). Below, in section~\ref{S:notion-motion}, we will employ a co-moving perspective to make this pseudo-Newtonian picture of $v^i$ clearer. A good, additional requirement is then to demand that this metric is \emph{not} static. This gives a better notion of movement, though it remains somewhat unclear if this asymptotic flatness and non-staticity are enough to exclude non-warp space-times.

Back to our main line of inquiry: From this line element one can easily read off a suitable orthonormal co-tetrad
\begin{equation}
	e^{\hat t}{}_a \; \d x^a = \d t; \qquad
	e^{\hat i}{}_a \; \d x^a = \d x^i - v^i \, \d t;
\end{equation}
and the corresponding orthonormal tetrad
\begin{equation}
	e_{\hat t}{}^a \; \partial_a = \partial_t + v^i\,\partial_i; \qquad
	e_{\hat i}{}^a \; \partial_a  = \partial_i.
\end{equation}
Viewed as $4\times4$ matrices these are simply
\begin{equation}
	e^{\hat b}{}_a = \left[ 
		\begin{array}{c|c}
			1 & 0\\  \hline\vphantom{\Big |} -v^{j} & \delta^{j}{}_{i}
		\end{array} \right]
	\qquad \hbox{and} \qquad
	e_{\hat b}{}^a = \left[ 
		\begin{array}{c|c}
			1 & + v^i\\ \hline 0  & \delta_{j}{}^{i}
		\end{array} \right],
\end{equation}
which satisfy the orthogonality conditions
\begin{equation}
	e^{\hat b}{}_a \; e_{\hat c}{}^a = \delta^{\hat b}{}_{\hat c};
	\qquad \hbox{and} \qquad
	e^{\hat b}{}_a \; e_{\hat b}{}^c = \delta_a{}^c.
\end{equation}
It is then a straightforward if somewhat tedious exercise to use the ADM formalism to calculate all of the orthonormal components of the Riemann, Ricci,  and Einstein tensors --- see reference~\cite{Santiago:2021a} for full details. 

A useful feature of these co-tetrads and tetrads is to note that the (space)--(space) portion is just the ordinary Kronecker delta, an observation which can be used to simplify the (space)--(space) portion of many calculations. For instance for any $T^0_2$ tensor $X_{ab}$ we have
\begin{equation}
	X_{\hat i \hat j} = e_{\hat i}{}^a \; e_{\hat j}{}^b \; X_{ab} =
	e_{\hat i}{}^j \; e_{\hat j}{}^k \; X_{jk} =  
	\delta_{i}{}^j \; \delta_{j}{}^k \; X_{jk} =  X_{ij}. 
\end{equation}
So for covariant spatial indices, within the Natario generic class of spacetimes,  one need not bother distinguishing orthonormal form Cartesian components. 

\subsection{Three specific simplified sub-classes}\label{S:classconsciousness}
\enlargethispage{20pt}

There are three major specific simplified sub-classes of the generic Natário class of warp drives that are of particular interest. 
\begin{description}
\item[Alcubierre:] 
For the Alcubierre warp bubble the flow field $\vec v(t,\vec x)$ is auto-parallel. That is 
\begin{equation}
	\vec v(t,\vec x) = v(t,\vec x)\; \hat v; \qquad 
	\hat v = \hbox{(constant)}.
\end{equation}
In this situation explicit computation shows that the NEC, and so the SEC, WEC and DEC,  are all explicitly violated.
See specifically the original discussion in~\cite{Alcubierre:1994}, and the subsequent follow-up discussions in~\cite{Visser:1999, Alcubierre:2001, Lobo:2002, Lobo:2004a, Lobo:2004b, Lobo:2007}, and~\cite{Lobo:2017, Alcubierre:2017}.
\item[Zero-expansion:] 
For the zero-expansion warp bubble the flow field $\vec v(t,\vec x)$ is taken to be solenoidal (divergence free). That is
\begin{equation}
	\nabla \cdot \vec v(t,\vec x) = 0.
\end{equation}
In this situation explicit computation shows that the NEC, and so the SEC, WEC and DEC,  are all explicitly violated.
See specifically the original discussion in~\cite{Natario:2001}, and subsequent follow-up discussions in~\cite{Lobo:2002, Lobo:2004a, Lobo:2004b, Lobo:2007}, 
and~\cite{Lobo:2017}.
\item[Zero-vorticity:]
For  the zero-vorticity warp bubble the flow field $\vec v(\vec t,x)$ is potential. That is~\cite{Lentz:2020, Lentz:2021,  Fell:2021},\:
\begin{equation}
	\vec v(t,\vec x) = \nabla \Phi(t,\vec x).
\end{equation}
In this situation the NEC, and so the SEC, WEC and DEC, energy conditions are still violated. 
\end{description}

Many of the early discussions of  energy condition violations in warp drive spacetimes often relied on either approximations or specific choices for the flow field to facilitate easy calculation. As said in the introduction, in previous work~\cite{Santiago:2021a}, the present authors provided a first, general proof of NEC violations of the \emph{generic} Natário warp drive. While the proof of NEC violations is fully general, it proceeds indirectly (a proof by contradiction) and hence implicitly---the proof cannot tell explicitly \emph{where} energy condition violations happen; just that they \emph{must} occur.

It is the zero-vorticity warp bubble that we shall focus on in the current article. Specifically, this is the form most appropriate to make the appearance of notions of mass in a warp drive explicit, as it can be easily compared to Schwarzschild in Painlevé--Gullstrand form. Additionally, we shall make use of its simplicity to highlight \emph{where} violations of the NEC occur, independent of mass. This should provide the added benefit of yielding better intuition for the general proof of \cite{Santiago:2021a}.

\subsection{ADM mass in warp drive spacetimes}
\label{SS:mass-drive}

Ever since their inception, a common side remark concerning warp-drive spacetimes was how the presence of a finite-mass space-ship (the payload) inside the warp bubble might or might not change the technical details. In this section we shall briefly explain why this particular concern is not really all that troublesome, and how non-zero masses actually \emph{can} play a role in warp drive metrics. At a minimum, there are three distinct ways in which mass can appear in a warp drive metric:
\begin{enumerate}
\item[A] As a mass (a payload, presumably a space-ship) 
inside the warp bubble.
\item[B] As the mass of the warp bubble itself (for sufficiently high mass this could be called a ``black warp'').
\item[C] As an external mass the warp drive is passing by.
\end{enumerate}
	
First of all, case A: Warp drives are a prime example of metric engineering, specifically \emph{reverse} engineering. As a reverse-engineered metric, it need not concern itself with possible interaction between its constituent parts: The metric is already fixed, essentially it is given \emph{by fiat}. To illustrate this, take any massive space-time metric $g_{{M}}$ and any warp metric $g_{\text{warp}}$. Suppose that the interior of the warp bubble has roughly radius $R$, and write the geodesic distance from the warp bubble's centre as $r$. Now, let $f_R(r)\geq 0$ be a smooth bump function centred on the middle of the warp bubble, such that $f_R(r)=0$ for any $r\geq R$, and $f_R(r)=1$ for some inner radius $R_0<R$. Any sufficiently advanced civilization interested in reverse-engineering spacetime metrics,  (not just in the sense of the \emph{purely mathematical}, technical meaning of ``{metric engineering}'' as it is used in this article!), could now easily engineer a new metric
\begin{equation}
        \label{metric}
	g \equiv (1-f_R(r)) \,g_{\text{warp}} + f_R(r)\, g_{{M}}.
\end{equation}
Inside the warp bubble, within the radius $R_0$, this will just be the metric of the mass inside. After a distance $R$ from the centre is reached, only the warp drive metric contributes. In between $R_0$ and $R$, the technology (indistinguishable from magic) of the arbitrarily advanced civilization will enforce a smooth transition between these two parts. Most importantly, the mass of the whole space-time will only depend on the warp bubble itself, as the new metric $g_{\text{warp}}$ effectively screens and possibly cancels completely the interior mass. In spirit, this may remind some readers of various proposals for regular black holes. Variations on this will be discussed in section~\ref{S:regular}.
	
This brings us to the second possibility, that of case B, that the warp bubble itself has a non-zero mass. This is most easily described in the 
Natário framework for generic warp bubbles \eqref{E:generic}, wherein the warp drive metric is not just locally in ADM form, but actually is also globally hyperbolic.
In this framework the ADM mass is defined by the large-distance asymptotic falloff in the flow vector
\begin{equation}
	v^i(t,\vec x) = v_0^i(t)+ \sqrt{ 2M\over r} \; \hat r^i + \O(r^{-3/2}).
\end{equation}	
Then this case B simply means that this warp metric has a finite ADM mass. For this reason, we will call this a ``Schwarzschild-based warp drive'' in section~\ref{S:black-warp} below. There are one quirky, and two important things to notice here. Starting with the important, moving to the quirky: 
	
\begin{itemize}
\item One, most extant warp drive metrics have ADM mass that is identically zero, as it keeps the headaches for interpreting the metric to a minimum. Specifically, both the Alcubierre and zero-expansion warp drives have identically zero ADM mass; whereas the zero-vorticity warp drive can, but need not, have a non-zero ADM mass. One does however want the ADM mass to be both non-negative and finite.  A negative ADM mass is observationally disfavoured, astronomers have looked for such objects and not seen them~\cite{Takahashi:2013}, and is also theoretically disfavoured~\cite{Bondi,Penrose:1993} --- causing  problems for both chronology protection and gravitational lensing; one would expect unusual caustics which do not seem to correspond to anything astronomers have ever seen~\cite{Cramer,Izumi:2013}. Infinite ADM masses are perhaps worse; completely undermining the notion of asymptotic flatness. Put differently, finiteness of the ADM mass places mild constraints on the fall-off of the metric components, for them not to be picked up by the integral in the equivalent formulation $M= \int \rho \, \d V$ of the ADM mass~\cite{Santiago:2021a}. 

\item	Two, as this discussion only depends on the asymptotic fall-off region of the warp bubble, this ADM mass is---according to the previous paragraph---entirely independent of whatever masses might be hidden inside the warp bubble.

\item Three, speaking about hiding: If the warp bubble's mass is significant enough to be hidden behind a horizon, one could properly call this a ``black-warp'' space-time. Here, it would be some kind of compact horizon buzzing (or crawling\dots) through space-time, reminiscent of the interpretation of the $C$-metric as an accelerated black hole---albeit with no strings attached \cite{CMetricInterpretation,GriffithsPodolsky}.
\end{itemize}\enlargethispage{20pt}		
Implementing option C can be significantly harder: Here, the idea is to let the warp drive move through another space-time that possibly has a finite mass $M$ of its own. The simplest example to imagine is a warp bubble moving \emph{outside} of, and manoeuvring around, a Schwarzschild black hole. (Especially a superluminal warp drive might have interesting, geometric things to say about how the compact black hole horizon and the non-compact warp drive horizon entwine; but that is another story and shall be told another time.) 

The most common examples, (albeit without a mass), are attempts (usually unsuccessful or otherwise doubtful for various reasons) to implement a warp bubble together with a cosmological constant. If we now focus on the mass aspects of this situation, it is easiest to interpret if the warp bubble by itself generates  no mass contributions. Essentially, this means that in this situation the warp drive metric is to be engineered as a zero ADM mass example of the second option, and then added to the ``{background}'' metric $g_M$ of ADM mass $M$ in a comparable but inverted way to the previously discussed option (\ref{metric}): 
\begin{equation}
	g \equiv (1-f_R(r)) \,g_{{M}} + f_R(r)\, g_{\text{warp}}.
\end{equation}
This would, however, be by far not the only possibility. Any less obvious change to the metric $g_M$ would require very careful thought as to whether or not other features, that should not change, might have changed.

If the warp drive part has a non-vanishing mass by itself, this will likely lead to subtle issues of disentangling the masses---a common issue with notions of mass in non-static situations. (Remember that the warp drive itself should be moving; the metric hence cannot be static!) This is not unique to this (rather unphysical) context. General relativity simply eschews straightforward implementations of notions of (total) mass and quasi-local mass based on the more familiar Newtonian gravity. 
(See for instance references~\cite{Wang:2020,Szabados:2009,Cederbaum}, and the textbook discussions in~\cite[chapter 19]{MTW} and~\cite[chapter 11.2]{Wald}.)
In particular, this should make us very worried if a ``{metric-engineered}'' warp drive (by itself or with a 
``background'' metric) has an undefined (vulgo: infinite) ADM mass. 

Lastly, it is obviously possible to combine the cases A to C in varied ways. As this only obfuscates otherwise easily discussed physics, we will opt \emph{not} to indulge in this needless complication.

\section{Zero-vorticity warp drives}
\label{S:zero-vorticity}

Let us now adapt and extend some of the discussion above, to refine the definition of a zero-vorticity warp bubble.

\subsection{Metric: The notion of motion}
\label{S:notion-motion}

For zero vorticity we can explicitly write the metric (line element) in the form:
\begin{equation}
	\d s^2 = -\d t^2 + \delta_{ij} \, 
	(\d x^i - \nabla^i\Phi(t,\vec x)\; \d t)\, (\d x^j - \nabla^j\Phi(t,\vec x)\; \d t), 
\end{equation}
Let us now extend the previous discussion, to make it more precise and manageable. Specifically, we will aim to divide the spacetime geometry  into a  ``base geometry'' and a ``warp field''. 

To make the warp bubble interesting, you want it to ``move''. 
The easiest way to do this is (we shall soon note at least one alternative) to ensure that $ \Phi(t,\vec x)$ really is time-dependent,
and the easiest way to ensure that is to enforce 
\begin{equation}
	\Phi(t,\vec x) \to \Phi\left(\vec x- \vec x_*(t)\right).
\end{equation}
That is, the zero-vorticity warp drive line element is taken to be 
\begin{equation}
	\d s^2 = -\d t^2 + \delta_{ij} \, 
	\left(\d x^i - \nabla^i\Phi(\vec x- \vec x_*(t) )\; \d t\right)\, 
	\left(\d x^j - \nabla^j\Phi(\vec x -\vec x_*(t)) \; \d t\right). 
\end{equation}
This is to be interpreted as follows: Start with some static asymptotically flat ``base'' spacetime geometry, described by the line element 
\begin{equation}
	(\d s_0)^2 = -\d t^2 + \delta_{ij} \, 
	\left(\d x^i - \nabla^i\Phi(\vec x) \; \d t\right)\, \left(\d x^j - \nabla^j\Phi(\vec x) \; \d t\right),
\end{equation}
where we choose $\vec v(\vec x) = \nabla\Phi(\vec x) \to 0$ at spatial infinity. We then
subject this ``base'' spacetime to a time-dependent spatial translation $\vec x_*(t)$, the ``warp field''. 
This is manifestly a specific example of a zero-vorticity warp drive, and it is important to note that the instances of vorticity-free warp drives given in~\cite{Lentz:2020, Lentz:2021,  Fell:2021} explicitly fall under this classification.
This particular version of the warp bubble has been carefully chosen to make the flow vector asymptote to zero at large distances --- so in this coordinate system the ``fixed stars'' are at rest, while the ``warp bubble'' is moving. 
\begin{equation}
	\d s^2 \to (\d s_\infty)^2 =  - \d \bar t^2 + \delta_{ij} \, 
	\d \bar x^i\, \d \bar x^j .
\end{equation}

Alternatively, one could choose coordinates $\bar x^a$ comoving with the warp bubble:
\begin{equation}
	t = t; \qquad \bar x^i = x^i - x_*^i(t); 
\end{equation}
so that
\begin{equation}
	\d t = \d t; \qquad \d \bar x^i = \d x^i - \dot x_*^i(t) \; \d t = \d x^i - 
	v_*^i(t) \; \d t.
\end{equation}
In these comoving coordinates the spacetime metric is 
\begin{equation}
	\d \bar s^2 = -\d  t^2 + \delta_{ij} \, 
	\left(\d \bar x^i - [\nabla^i\Phi(\bar{x}^k) +v_*^i(t)] \; \d t\right)\, 
	\left(\d \bar x^j - [\nabla^j\Phi(\bar{x}^k )+v_*^j(t)]\; \d t\right).
\end{equation}
At spatial infinity, where $v^i(\bar x^k)= \nabla^i\Phi(\bar x^k)\to 0$, we see that
\begin{equation}
	\d \bar s^2 \to (\d \bar s_\infty)^2 =  -\d \bar t^2 + \delta_{ij} \, 
	\left(\d \bar x^i - v_*^i(t) \; \d t\right)\, 
	\left(\d \bar x^j - v_*^j(t)\; \d t\right).
\end{equation}
So in these coordinates it is the ``warp bubble'' that is ``at rest'', while the ``fixed stars'' are moving. 

Either one of these two coordinate equivalent spacetime line elements, 
\begin{align}
	\d s^2 &= -\d t^2 + \delta_{ij} \, 
	\left(\d x^i - \nabla^i\Phi(\vec x- \vec x_*(t) )\; \d t\right)\, 
	\left(\d x^j - \nabla^j\Phi(\vec x -\vec x_*(t)) \; \d t\right),
	\\[7pt]
	\d \bar s^2 &= -\d  t^2 + \delta_{ij} \, 
	\left(\d \bar x^i - [\nabla^i\Phi(\bar{x}^k) +v_*^i(t)] \; \d t\right)\, 
	\left(\d \bar x^j - [\nabla^j\Phi(\bar{x}^k )+v_*^j(t)]\; \d  t\right),
\end{align}
represents exactly the same spacetime physics, and 
they are  equally valid ways of representing the warp bubble --- which one you use is a matter of choice --- but however one does it, one needs either the explicit spatial shift $x_*^i(t)$ or the warp flow $v_*^i(t)=\dot x_*^i(t) $ to encode the ``notion of motion'' of the warp bubble with respect to the fixed stars. 

In either situation, either by setting the spatial shift $x_*^i(t)\to0$,
 or by setting the 
flow $v_*^i(t)\to0$, one recovers the same ``base'' spacetime geometry
\begin{equation}
	(\d s_0)^2 = \d t^2 + \delta_{ij} \, 
	\left(\d x^i - \nabla^i\Phi(\vec x) \; \d t\right)\, 
	\left(\d x^j - \nabla^j\Phi(\vec x) \; \d t\right).
\end{equation}
(For the base geometry, since one has switched off the warp bubble,  one does not need to distinguish $x^i$ from $\bar x^i$.)

Now that we have refined the zero-vorticity notion of warp drive by introducing these notions of  ``base'' and ``warp'', we can ask more precisely targeted question such as this: ``How precisely does the stress-energy tensor \emph{change} when you switch the  warp field on or off?''.
Fortunately all of the relevant tools have been developed in earlier work~\cite{Santiago:2021a}.

\subsection{Einstein tensor and stress-energy tensor}
\label{SS:Einstein-tensor}
\def\L{{\mathcal{L}}}

For the generic Narari\'o warp drive, and so also for the zero-vorticity warp drive,  we have previously determined the full stress-energy tensor in~\cite{Santiago:2021a}. Let us now split the zero-vorticity warp bubble into  ``base'' and ``warp'' and adopt comoving coordinates, dropping the over-bar on the $\bar x^i$ for brevity,  so that
\begin{align}
	\d  s^2 &= -\d t^2 + \delta_{ij} \, 
	\left(\d  x^i - [v^i({x}^k) +v_*^i(t)] \; \d t\right)\, 
	\left(\d x^j - [v^j({x}^k )+v_*^j(t)]\; \d t\right). 
	\\[7pt]
	(\d s_0)^2 &= -\d t^2 + \delta_{ij} \, 
	\left(\d x^i - v^i(\vec x) \; \d t\right)\, \left(\d x^j - v^j(\vec x) \; \d t\right).
\end{align}
For the extrinsic curvature $K_{ij} = v_{(i,j)}$, using the fact that the warp field $v_*^k(t)$ is spatially constant, $\partial_i v_*^k(t)=0$,  we have the particularly simple result that
\begin{equation}
	K_{ij} = (K_0)_{ij}.
\end{equation}
In these comoving coordinates, the extrinsic curvature does not change as you switch on the warp field.
In contrast, for the Eulerian 4-velocity we have
\begin{equation}
	n^a = \left( 1; v^i({x}^k) +v_*^i(t) \right) = (n_0)^a + \left( 0; v_*^i(t) \right).
\end{equation}
This is a simple linear sum of the base spacetime contribution and the warp field contribution. This in turn affects the Lie derivatives of the extrinsic curvature. Specifically 
\begin{equation}
	\L_n K = \L_{n_0} K + v_*^i(t) \; \partial_i K,
\end{equation}
and
\begin{equation}
	\L_n K_{ij} = \L_{n_0} K_{ij} + v_*^k(t) \; \partial_k K_{ij}.
\end{equation}
Feeding this into the Einstein tensor, for the tetrad components we find 
\begin{equation}
	G_{nn} = (G_0)_{nn}; \qquad G_{ni} = (G_0)_{ni}; \qquad
	G_{ij} = (G_0)_{ij} + v_*^k(t) \; \partial_k [K_{ij} - K \delta_{ij}].
\end{equation}
Applying the Einstein equations, for the tetrad components one has 
\begin{equation}
	\rho = \rho_0; \qquad f_{i} = (f_0)_{i}; \qquad
	T_{ij} = (T_0)_{ij} + {1\over 8\pi} \; v_*^k(t) \; \partial_k [K_{ij} - K \delta_{ij}].
\end{equation}
So we see that it is only the spatial parts of the stress-energy that are affected by switching on the warp field --- and even then the effect is rather simple --- a contribution \emph{linear} in the warp field $v_*^k(t)$. 

If we explicitly make use of the zero-vorticity condition then one sees 
\begin{equation}
	\rho = \rho_0; \qquad f_{i} = 0; \qquad
	T_{ij} = (T_0)_{ij} + {1\over 8\pi} \; v_*^k(t) \; \partial_k [\Phi_{,ij} - \nabla^2\Phi\, \delta_{ij}].
\end{equation}

Now consider the average pressure $\bar p = {1\over 3} \delta^{\hat i\hat j}\, T_{\hat i\hat j} = {1\over 3} \delta^{ij}\, T_{ij}$. From the above we have
\begin{equation}
	\bar p = \bar p_0 - {1\over 12\pi} \; v_*^k(t) \; \partial_k \nabla^2\Phi.
\end{equation}
We can rewrite this as 
\begin{equation}
	\bar p =  \bar p_0 - {1\over 12\pi} \; \vec v_* \cdot  \nabla (\nabla^2\Phi)
\end{equation}
Furthermore for the quantity $(\rho+ \bar p)$ that is of direct relevance to testing the NEC 
\begin{equation}
	(\rho+ \bar p) = (\rho_0 + \bar p_0) - {1\over 12\pi} \; \vec v_* \cdot  \nabla (\nabla^2\Phi)
\end{equation}
Note the effect of switching on the warp bubble is \emph{linear} in the warp field $v_*^k(t)$.

\section{Schwarzschild-based warp drive}
\label{S:black-warp}

Now, using the framework developed above,  let us develop the notion of a Schwarzschild-based warp drive---the above option B that might at times be called a black-warp space-time.

\subsection{Schwarzschild spacetime in Painlev\'e--Gullstrand form}
\label{SS:Schwarzschild-PG}

It is well-known that Schwarzschild spacetime can be represented in Painlev\'e--Gullstand form as follows:
\begin{equation}
	\d s^2 = - \d t^2 + \delta_{ij} 
	\left(\d x^i - \sqrt{2m/r}\; \hat r^i\; \d t\right)\, 
	\left(\d x^j  -\sqrt{2m/r} \; \hat r^j\; \d t\right),
\end{equation}
Here $r= |\vec x| = \sqrt{x^2+y^2+z^2}$, and  $\hat r^i $ is the radial unit vector: $\hat r^i = {x^i\over r} = {x^i\over|\vec x|}$. 

This is fully equivalent to writing
\begin{equation}
	\d s^2 = - \d t^2 + \delta_{ij} 
	\left(\d x^i - \sqrt{2m}\; {x^i\over|\vec x|^{3/2}}\; \d t\right)\, 
	\left(\d x^j  -\sqrt{2m} \; {x^j\over|\vec x|^{3/2}}\; \d t\right).
\end{equation}
When written in this manner the Schwarzschild spacetime has a number of interesting features, including the fact that these coordinates are horizon-penetrating, and that the so-called ``drip'' geodesics (corresponding to radially infalling geodesics that start from spatial infinity with zero 3-velocity) are particularly simple. (See for instance references~\cite{unit-lapse,PGLT,carter}.)
The relevant potential $\Phi$ is easily seen to be 
\begin{equation}
	\Phi(\vec x) = 2 \sqrt{2mr} = 2 \sqrt{2 m\, ||\vec x||}. 
\end{equation}
(This is \emph{not} the usual Newtonian potential.)

\subsection{Line element for Schwarzschild-based warp drive}
\label{SS:black-warp-line-element}

Elevating the  Painlev\'e--Gullstrand form of Schwarzschild to a Schwarzschild-based warp drive (black-warp spacetime) merely amounts to the replacement $\vec x \to \vec x - \vec x_*(t)$ so that the spacetime metric becomes:
\begin{equation}
	\d s^2 = - \d t^2 + \delta_{ij} 
	\left(\d x^i - \sqrt{2m}\; {x^i-x_*^i(t) \over|\vec x- \vec x_*(t)|^{3/2}}\; \d t\right)\, 
	\left(\d x^j  -\sqrt{2m} \; {x^j -x_*^i(t)\over|\vec x- \vec x_*(t)|^{3/2}}\; \d t\right).
\end{equation}
In this coordinate system the ``fixed stars'' are ``at rest''  and the warp bubble is ``moving''. At large distances
\begin{equation}
	\d s^2 = - \d t^2 + \delta_{ij} \, \d x^i\, \d x^j.
\end{equation}

If we change to comoving coordinates $\bar x^i = x^i-x_*^i(t) $ then by our previous arguments the black-warp line element becomes
\begin{equation}
	\d \bar s^2 = - \d t^2 + \delta_{ij} 
	\left(\d \bar x^i - 
	\left[\sqrt{2m}\; {\bar x^i \over|\vec{\bar x}|^{3/2}}+ v_*^i(t)\right]
	\; \d t\right)\, 
	\left(\d \bar x^j  -
	\left[\sqrt{2m}\; {\bar x^j \over|\vec{\bar x}|^{3/2}}+ v_*^j(t)\right]
	\; \d t\right).
\end{equation}
In this coordinate system the ``fixed stars'' are ``moving''  and the warp bubble is ``at rest''.
At large distances one now has
\begin{equation}
	(\d \bar s_\infty)^2 = - \d t^2 + \delta_{ij} 
	\left(\d \bar x^i - v_*^i(t)\; \d t\right)\, 
	\left(\d \bar x^j  - v_*^j(t)\; \d t\right).
\end{equation}

\subsection{Energy conditions for Schwarzschild-based warp drive}
\label{SS:black-warp-ECs}
In view of the fact that the ``base'' stress-energy is zero for Schwarzschild spacetime, in coordinates moving with the Schwarzschild-based warp drive we have the very simple results
\begin{equation}
	\rho = 0; \qquad f_{i} = 0; \qquad
	T_{ij} =  {1\over 8\pi} \; v_*^k(t) \; \partial_k [\Phi_{,ij} - \nabla^2\Phi\, \delta_{ij}];
	\qquad
	\Phi = 2\sqrt{2mr}.
\end{equation}
Without any calculation, since $\rho=0$, while $T_{ij}\neq 0$, we immediately deduce DEC violations.

Furthermore taking the spatial trace
\begin{equation}
	\bar p = {1\over 3} \,\delta^{\hat i\hat j} \,T_{\hat i\hat j} =
	{1\over 3} \,\delta^{ij} \,T_{ij} =
	-{1\over 12\pi} \; v_*^k(t) \; \partial_k  \nabla^2\Phi.
\end{equation}
Note that
\begin{equation}
	\nabla^2 \sqrt{r} = \nabla\cdot\left({1\over2}\, r^{-1/2} \,\hat r \right) =
	\nabla\cdot\left({1\over2} \,r^{-3/2}\, \vec r \right) = {3\over4} \,r^{-3/2},
\end{equation}
and consequently
\begin{equation}
	\partial_k  \nabla^2\Phi = -{9\over 4} \sqrt{2m} \; r^{-5/2} \; (\hat r)_k
\end{equation}
Thence
\begin{equation}
	\bar p = 
	{3\over 16\pi} \; {\sqrt{2mr}\over r^3} \; \left\{ \vec v_*(t) \cdot  \hat r \right\}.
\end{equation}
But the key observation here is that the inner product $\left\{\vec v_*(t) \cdot  \hat r\right\}$ changes sign as one moves from front to back of the warp bubble; therefore there are regions (an entire hemisphere in fact) where $\bar p$ is negative. Since $\rho=0$ by construction, there is consequently an entire hemisphere where $\rho+\bar p <0$ and the NEC is explicitly violated. 
(So, since DEC $\implies$ WEC $\implies$ NEC, and SEC $\implies$ NEC, the explicit violation of the NEC implies that  all of the SEC, WEC, and DEC are also violated.)

\section{Regular warp bubble containing a massive payload}
\label{S:regular}

Let us now modify the Schwarzschild-based warp drive (black-warp spacetime), to bring it more into line with what we would expect a warp bubble containing a payload (spacecraft) to look like. As discussed in section~\ref{S:classconsciousness}, we can expect this to have surprisingly little impact on the physical situation based on general arguments. However, there is nothing like practice to understand something, so let us re-examine the NEC in this context once more. Having done so, we can then also slightly modify and further generalize the argument.

\subsection{Localized regular base spacetime}
\label{SS:localized}

To avoid horizons and singularities, and more closely envisage the notion of a spaceship embedded in a warp bubble,  one can modify the base spacetime by making it regular at short distances and Schwarzschild at large distances. Specifically choose some finite radius $a > 2m$ and set:
\begin{equation}
	\Phi(r) = \left\{ 
		\begin{array}{cc} 
			\O(r^2) & r<a;\\[7pt]
			\hbox{\;\;differentiable\;\;} & r=a; \\[7pt]
			2\sqrt{2mr} & r > a.
		\end{array}\right._.
\end{equation}
The $\Phi(r) = \O(r^2) $ condition keeps the curvature and stress-energy finite at the origin. (The location of the spacecraft.) 
The differentiability condition at $r=a$ keeps $\nabla\Phi$ continuous, and so keeps the metric continuous. 
(Hence the Christoffel symbols are at worst discontinuous and the Riemann tensor at worst contains thin-shell delta functions.) 
For $r>a$ one has a vacuum Schwarzschild exterior.

The argument presented above for the Schwarzschild-based warp drive (black-warp spacetime) then guarantees (once one switches on a non-zero warp velocity $\vec v_*$) violation of the NEC in the exterior region $r>a$ over the  entire hemisphere where $\left\{\vec v_*\cdot \hat r\right\} <0$.

\subsection[Asymptotically Schwarzschild base spacetime]{
Asymptotically Schwarzschild base spacetime}
\label{SS:asymptotic}
One can also extend the argument to a base spacetime that is only asymptotically Schwarzschild. Specifically let us set
\begin{equation}
	\Phi(r) = \left\{ 
		\begin{array}{cl} 
			\O(r^2), &\quad r\to0;\\[7pt]
			\hbox{\;\;differentiable,\;\;} &\quad \hbox{all $r$}; \\[7pt]
			2\sqrt{2mr}\;[1 +\O(r^{-n})],& \quad r \to\infty, \;\; n > 0. 
		\end{array}\right._.
\end{equation}
The $\Phi(r) = \O(r^2)$ condition keeps the curvature and stress-energy finite at the origin. 
The differentiability condition at all $r$ keeps $\nabla\Phi$ continuous, and so keeps the metric continuous. 
For $r\to\infty$ one asymptotically approaches a vacuum Schwarzschild exterior with Misner--Sharp quasi-local mass $m(r) = m\;\left[1 +\O(r^{-n})\right]$ while the ADM mass is simply $m$.
Then it is easy to check that
\begin{equation}
	\partial_k  \nabla^2\Phi = -{9\over 4} \sqrt{2m} \, r^{-5/2}\;
	 (\hat r)_k \;\left[1 +\O(r^{-n})\right],
\end{equation}
and consequently 
\begin{equation}
	\bar p = 
	{3\over 16\pi} \; {\sqrt{2mr}\over r^3} \; [1 +\O(r^{-n})]\; 
	\left\{ \vec v_*(t) \cdot  \hat r\right\}.
\end{equation}
Thence at sufficiently large distances there is again an entire hemisphere, (defined by $ \left\{\vec v_*(t) \cdot  \hat r \right\} <0$ ), where $\bar p<0$. 
So the NEC, (and consequently all of the SEC, WEC, and DEC), are again violated at asymptotically large distances. 

\section{Conclusions}
\label{S:conclusions}

As we have seen, mass in warp drives is more than just a question of a ``payload''---there are two more cases to consider: The possibilities of a ``background'' with mass or the possibility the ``warp bubble'' itself being massive. Our guide for this discussion was that the assumed asymptotic flatness of a warp drive allows invoking the concept of an ADM mass, and hence, to studying how the ADM mass influences and is influenced by the presence of a warp drive.
	
Additionally, we re-investigated the status of energy conditions in these space-times. Certainly, the status of the energy conditions in all of these classes has already been covered in the fully general arguments presented in the authors' previous study~\cite{Santiago:2021a}. Still, the present article can \emph{explicitly} show \emph{where} violations of the NEC (and hence all other common energy conditions) occur. This was possible by specializing to the case of zero-vorticity Natário warp drive with sufficient emphasis on how the mass can enter. Given that proofs of energy condition violations often involve implicit or indirect arguments (like the proofs by contradiction), this provides additional, helpful intuition about their location and these decades-old results and convictions concerning warp drives. And as the discussion in the literature amply shows, in the case of warp drives (particularly those of the superluminal variety) their violations of energy conditions \emph{are} a sign of ``bad physics''---though these violations themselves by themselves are at best a warning sign \cite{twilight,Curiel}.

For the future one could try to develop further generalizations of the notion of warp bubble, outside of the Natário generic class. This could be done  either by relaxing the unit-lapse condition,  or by allowing the spatial 3-slices to deviate from being geometrically flat. (Perhaps conformally 3-flat.) The ``massive background'' case discussed here can certainly be considered as first step in this direction. However, there is a crucial and unavoidable trade-off between tractability and generality. If the construction is too general then not only are stress-energy computations increasingly infeasible~\cite{Santiago:2021a}, but it also becomes much trickier to even define what one means by the ``warp bubble'', and how to disentangle the ``warp field'' from the ``payload'' from the ``rest of the universe''. We hope to further explore such issues in future work. Future work will also study how the (non-compact) horizon of a superluminal warp drive might interact with the compact horizon of a black hole either moving as a warp drive, inside a warp drive as payload, or present as an external immovable background geometry.

Lastly, a word on the most glaring, open problem: A self-consistent formulation of warp drives. Simply calculating the stress-energy tensor by reverse-engineering the Einstein equations is telling us little about the actual matter sourcing the warp drive. We \emph{know} that a more realistic scenario would entail reverse-engineering at the very least something like Einstein--Klein--Gordon or Einstein--Maxwell. With a proverbial ``here be dragons'', one could go a step further and look at quantized fields on a fixed background. However, the background being fixed, this is unlikely to answer how the quantized field is actually sourcing anything if it is not yet a source term of the Einstein equation determining the warp drive metric. An added difficulty is that a superluminal warp drive comfortably overstays its welcome in the well-established field of curved space-time quantum field theory, as it cannot be globally hyperbolic \cite{Finazzi:2009}. While there is movement beyond global hyperbolicity (see, for example, \cite{Janssen2022}), one could still opt for the full glory: Dealing with the full semi-classical Einstein equations involving a quantized field on a (backreacting) background space-time \cite{MPS2020,Sanders2022}. While this would certainly (finally) address and answer the begged question of the usual invocation of arbitrarily advanced civilizations/magic by nebulous statements such as ``the negative energy densities would have to be provided by quantum matter''---here be dragons, indeed.

\section*{Acknowledgements}

MV was directly supported by the Marsden Fund, \emph{via} a grant administered by the Royal Society of New Zealand. \\
SS acknowledges support from the technical and administrative staff at the Charles University, and financial support from Czech Science Foundation grant No 22-14791S.\\
JS was supported by the Hellenic Foundation for Research and Innovation (H.F.R.I.),
under the “First Call for H.F.R.I. Research Projects to support Faculty members and Researchers and the procurement of high-cost research equipment grant”. (Project Number: 789)


\bigskip
\bigskip
\hrule\hrule\hrule
\enlargethispage{20pt}

\end{document}